\documentstyle[11pt]{article}
\normalbaselines
\begin{document}
\begin{center}{\large \bf
UNIVERSAL NEWTON TIME IN CLASSICAL ELECTRODYNAMICS.
ELEMENTS OF PHYSICAL INTERPRETATION}\\[2mm]
G.A. Kotel'nikov\\{\it Russian Research Center "Kurchatov Institute",
Moscow 123182, Russia}\\
E-mail: kga@kga.kiae.su\\[3mm]
\end{center}
\begin{abstract}
    It is  shown  that  the universal Newton time may be introduced in
the classical electrodynamics. The statement results from an existence
of  the  generalized  symmetry  of  Maxwell  equations with respect to
Galilei  transformations.  In  the  case  of  the   extended   Galilei
transformations  the postulate of invariance of the speed of light may
be made compatible with the concept of the universal Newton time. Some
physical consequences of the extended Galilei symmetry are considered.
\end{abstract}

\section{Introduction}
\label{1}
\par The postulate on the existence of the universal time was  entered
into  physics  by  Newton in 1687.  Up to the end of the XIXth century
this concept perfectly corresponded to the known set  of  experimental
facts, described  by  the  equations  of  movement  in  the  classical
mechanics both on the Earth and within the Sun system. It will suffice
to mention one of the remarkable achievements of this concept, namely,
the prediction of the Neptune's orbit parameters on the  basis  of the
Uran's perturbation study by Le Verrier,  1846.  The difficulties have
begun at the end of the  XIXth  century  in  attempting  to  make  the
universal time concept compatible with Maxwell electrodynamics.  It is
known from a formally mathematical side that the idea of the universal
time is realized in Galilei transformations:
\begin{equation}
\label{f1}
x' = x - Vt; \ y' = y; \ z' = z; \ t' = t.
\end{equation}
Here $x,  \ y,  \ z$ are the space variables,  $t$ is the time,  ${\bf
V}=(V,0,0)$ is the velocity of the inertial reference $K'$ relative to
$K$.  While the Newton mechanics equation $m{\bf  a}={\bf  F}$  is  in
accordance  with  these  transformations  (invariant  with  respect to
Galilei transformations),  Maxwell equations did not seem to have  the
similar   property.  In  other  words,  Maxwell  equations,  by  their
mathematical nature were presented as Galilei non-invariant equations.
This point of view is popularized everywhere. One can find this in any
university course of general physics.  Moreover,  it is from resolving
the difficulty of Galilei approach in electrodynamics that the special
relativity theory (SR) and the  modern  relativization  of  space-time
representations  have arisen.  And now about 90 years after publishing
the  fundamental  works  of  Voigt,  Larmor,  Lorentz,  Einstein   and
Poincar\'e we return to seemingly already resolved problems again.
\par It was found that a situation is more complicated here,  than  it
is  commonly supposed,  and the concept of the Newton time may be also
accorded with Maxwell equations.  It is important to determine what we
mean  by  this.  For  definiteness,  by  the words "the concept of the
Newton time may be also accorded with Maxwell equations" we shall mean
that vacuum  (microscopic)  Maxwell  equations  possess  not  only the
property of the Lorentz invariance,  but  also  the  property  of  the
invariance with respect to Galilei transformations in some generalized
sense. The Galilei transformations contain the time transformation $t'
=   t$,  and  the  universal  Newton  time  will  be  thereby  entered
automatically into Maxwell equations.
\par It should be noted that although  the  relativistic  concept  was
dominant,  the research of Galilei symmetry of Maxwell electrodynamics
were carried out after arising SR and are  continued  at  the  present
time. As  example it is possible to note the publications of Le Bellac
and Levi-Leblond \cite{Bel73} and Fushchich and Nikitin  \cite{Fus83}.
The  authors have established that besides the well known relativistic
and conform symmetry,  the first and second pair of Maxwell equations,
considered  separately,  have  also  the  property  of invariance with
respect to Galilei transformations. The various linear representations
of Galilei group corresponding to the transformations  of  the  fields
${\bf E}'={\bf E}$, ${\bf H}' =({\bf H}-{\bf V}$ x ${\bf E})$; \ ${\bf
E}'= ({\bf E} + {\bf V}$x${\bf H})$,  ${\bf H}'={\bf H}$ are  realized
on the solutions of these equations \cite{Bel73},  \cite{Fus83}. (Here
the speed of light $c$ is believed equal to unit; \ ${\bf E}, {\bf H}$
are  the  electrical  and  magnetic  fields).  So,  the  situation has
considerably  been  cleared,  but  the  attempt  at  establishing  the
property of Galilei symmetry of Maxwell equations in the total set has
not been quite successful. It was made later in the works \cite{Kot85}
- \cite{Kot91}, which we shall follow below.

\section{Galilei symmetry of Maxwell equations}
\label{2}
Let us introduce the free Maxwell equations:
\begin{equation}
\label{f2}
\begin{array}{cc}
\vspace{2mm}
\displaystyle
\nabla \cdot {\bf E}=0; \ \nabla \cdot {\bf H}=
+\frac {1}{c} \partial_t{\bf E};                                    \\
\displaystyle
\nabla \cdot {\bf H}=0; \ \nabla \cdot {\bf E}=
-\frac {1}{c} \partial_t{\bf H}.
\end{array}
\end{equation}
As it is known, they may be brought to a single equation of the second
order namely to the D'Alembert equation with $\phi \in ({\bf E},  {\bf
H})$:
\begin{equation}
\label{f3}
\vspace{2mm}
\Box \phi(x)=
\biggl(\frac{1}{c^2}{\partial^2}_{tt} -\triangle\biggr)\phi(x)=0.
\end{equation}                                                       
The algebraic   equation   of   the  light  cone  may  be  brought  in
correspondence with the differential D'Alembert equation:
\begin{equation}
\label{f4}
c^2t^2 - {\bf x}^2 =0.
\end{equation}                                                       
Any university course of general physics contains the statement on its
Galilei non-invariance.  Let us show by direct calculation that it  is
not  so.  For  this  purpose we use the transformations (\ref{f1}) and
Galilei theorem of velocities addition,  recorded  for  the  speed  of
light:
\begin{equation}
\label{f5}
c'_x = c_x -V = n_x c - V; \ c'_y = c_y = n_y c; \ c'_z =c_z = n_z c.
\end{equation}
Let us put (\ref{f1}) and (\ref{f5}) into the equation of light
cone in frame Š'. We have:
\begin{equation}
\label{f6}
\begin{array}{c}
\vspace{2mm}
\displaystyle
c'^2 t'^2-{\bf x}'^2=0 \to                                          \\
\vspace{2mm}
\displaystyle
({c'_x}^2+{c'_y}^2+{c'_z}^2)t'^2-{\bf x}'^2=
{(n_x c-V)}^2t^2+{n_y}^2c^2 t^2+{n_z}^2c^2 t^2-
{(x-Vt)}^2-y^2 -z^2=                                                \\
\vspace{2mm}
\displaystyle
({n_x}^2+{n_y}^2+{n_z}^2)c^2 t^2-
2n_x c V t^2+V^2 t^2-x^2-y^2-z^2+2xVt-V^2t^2=
c^2 t^2-{\bf x}^2=0.
\end{array}
\end{equation}                                                       
Here it  is  taken  into account that $x=c_xt=cn_xt$.  It follows from
here  that  the  Galilei  transformations   map   the   sphere   ${\bf
x}'^2=c'^2t'^2$  with center in the point $x'=y'=z'=0$ onto the sphere
${\bf x}^2=c^2t^2$ with the center in the point $x=y=z=0$ just as  the
Lorentz transformations,  which accords with the relativity principle.
The failure in the former considerations consists in the  fact  that
space  and time were transformed according to Galilei but the speed of
light was  transformed  according  to Lorentz.  Such inconsistency has
resulted to incorrect conclusion  that  the  relativity  principle  is
violated   at   joint  consideration  of  the  ratios  (\ref{f1})  and
(\ref{f4}).  In the fact the light cone  equation  is  invariant  with
respect  to  Galilei  transformations.  This  circumstance  prompts to
consider  the  symmetric  properties  of  D'Alembert   equation   more
carefully.

     In accordance  with  theoretical-algebraic  definition   of   the
symmetry \cite{Lez86}, let us consider the commutational ratios of the
Lie algebra generators of Galilei group with the D'Alembert  operator.
We have \cite{Kot85}:
\begin{equation}                                                       
\label{f7}
[\Box, p_0]=[\Box, p_k]=[\Box, J_k]=0; \ [\Box[\Box, {\it H}_k]=0.
\end{equation}
Here $p_0=i\partial_t$,   $p_k=-i\partial_k$   $J_k={(\bf   x}$x${{\bf
p})}_k$,  ${\it H}_k=-tp_k$,  $k=1,2,3$, $x_{1,2,3}=x,y,z$. It follows
from here that the Lie algebra of  Galilei  group  is  the  invariance
algebra  of  the  free  Maxwell  equations.  Hence,  the  set of field
transformations should exist which will transform into themselves  the
equations  (\ref{f2})  in combination with the Galilei transformations
(\ref{f1}).  The task is to find them.  For this we record the  sought
transformations as:
\begin{equation}
\label{f8}
\begin{array}{ll}
\vspace{2mm}
\displaystyle
E_1'=\Phi(x,t)E_1;              &  H_1'=\Phi(x,t)H_1;               \\
\vspace{2mm}
\displaystyle
E_2'=\Phi(x,t)a(E_2+h_{23}H_3); & H_2'=\Phi(x,t)a(H_2+e_{23}E_3);   \\
\vspace{2mm}
\displaystyle
E_3'=\Phi(x,t)a(E_3+h_{32}H_2); & H_3'=\Phi(x,t)a(H_3+e_{32}E_2),
\end{array}
\end{equation}                                                         
where $\Phi(x,t)$ is some weight function.  Let us express the Galilei
transformations in variables $x_0=ct$, ${\bf x}=(x_1,x_2,x_3)$:
\begin{equation}
\label{f9}
{x_0}'=\gamma x_0; \ {x_1}'=x_1-\beta x_0; \ {x_2}'=x_2; \ {x_3}'=x_3;
\ c'=\gamma c,
\end{equation}                                           
where $\gamma=(1-2\beta n_1+\beta^2)^{1/2}$, ${\bf n}={\bf c}/c= (n_1,
n_2,n_3)$,  $\beta=V/c$,  and produce the replacement of the variables
in  the  equations  (\ref{f2}),  \  (\ref{f3}) taking into account the
relations (\ref{f8}) and the relationships between the derivatives:
\begin{equation}
\label{f10}
\begin{array}{c}
\vspace{2mm}
{\partial_0}'=(\partial_0 + \beta \partial_1)/\gamma; \ {\partial_k}'=
\partial_k; \ k=1,2,3;                                              \\
\vspace{2mm}
{\partial_{00}}'^2={(\partial_0 + \beta\partial_1)}^2/\gamma^2=     \\
({\partial_{00}}^2+2\beta\partial_0\partial_1+\beta^2{\partial_{11}}^2)
/\gamma^2; \ {\partial_{kk}}'^2={\partial_{kk}}^2.
\end{array}
\end{equation}
Let us  use  the  concept of the generalized symmetry \cite{Kot96} and
require  that  the  initial  equations  should  be  transformed   into
themselves  due  to the compatibility of the set of engaging equations
(the conditions  of  transformation  into  themselves  and  the  final
equations), where the conditions of transformation into themselves are
obtained by means of  replacement  of  the  variables  (\ref{f8})  and
(\ref{f10}) in Maxwell equations (\ref{f2}).
\begin{equation}
\label{condition}
\begin{array}{ll}
\vspace{2mm}
{\bf Initial \ Equations}                                          &
{\bf Conditions \ of \ Transformation \ into \
Themselves. \ Final \ Equations}                                   \\
\nabla\cdot{\bf E}'=0;                                             &
\partial_1\Phi E_1+a\partial_2 \Phi (E_2+h_{23}H_3)+
a\partial_3 \Phi (E_3+h_{32}H_2)=0;                                \\
\vspace{1mm}
\nabla\cdot{\bf H}'=0;                                             &
\partial_1\Phi H_1+a\partial_2 \Phi (H_2+e_{23}E_3)+
a\partial_3 \Phi (H_3+e_{32}E_2)=0;                                \\
\nabla {\rm X}{\bf H}'-\partial'_0 {\bf E}'=0;                     &
a\partial_2\Phi (H_3+e_{32}E_2)-a\partial_3\Phi (H_2+e_{23}E_3)-
(\partial_0+\beta\partial_1)\Phi E_1/\gamma=0;                     \\
{} & \partial_3\Phi H_1-a\partial_1\Phi (H_3+e_{32}E_2)-
     a(\partial_0+\beta\partial_1)\Phi (E_2+h_{23}H_3)/\gamma=0;   \\
\vspace{1mm}
{} & a\partial_1\Phi (H_2+e_{23}E_3)-\partial_2\Phi H_1-
     a(\partial_0+\beta\partial_1)\Phi (E_3+h_{32}H_2)/\gamma=0;   \\
\nabla {\rm x}{\bf E}'+\partial'_0 {\bf H}'=0;                     &
a\partial_2\Phi (E_3+h_{32}H_2)-a\partial_3\Phi (E_2+h_{23}H_3)+
     (\partial_0+\beta\partial_1)\Phi H_1/\gamma=0;                \\
{} & \partial_3\Phi E_1-a\partial_1\Phi (E_3+h_{32}H_2)+
     a(\partial_0+\beta\partial_1)\Phi (H_2+e_{23}E_3)/\gamma=0;   \\
\vspace{2mm}
{} & a\partial_1\Phi (E_2+h_{23}H_3)-\partial_2\Phi E_1+
     a(\partial_0+\beta\partial_1)\Phi (H_3+e_{32}E_2)/\gamma=0;   \\
{} & \nabla\cdot{\bf E}=0;                                         \\
{} & \nabla\cdot{\bf H}=0;                                         \\
{} & \nabla {\rm X}{\bf H}-\partial_0 {\bf E}=0;                   \\
{} & \nabla {\rm X}{\bf E}+\partial_0 {\bf H}=0.
\end{array}
\end{equation}
The weight function $\Phi$ entering into this set may  be  found  from
the  condition  of  transformation of D'Alembert equation into himself
\cite{Kot85}:
\begin{equation}
\label{f11}
\begin{array}{ll}
\vspace{2mm}
{\bf Initial \ Equation} & {\bf Condition \ of \ Transformation \ into
\ Himself. \  Final \ Equation}                                         \\
\Box\phi'=0                                                             &
[{(\partial_0 + \beta\partial_1)}^2/\gamma^2 - \triangle] \Phi \phi=0;  \\
{}                       & \Box\phi=0.
\end{array}
\end{equation}
For  the  plane waves
\begin{equation}
\label{f12}
\begin{array}{cc}
\displaystyle                                                      
{\bf E}={\bf l} \ e^{-ik\cdot x }; & {\bf H}={\bf m} \ e^{-ik\cdot x},
\end{array}
\end{equation}
where ${\bf l}$  and  ${\bf m}$ are the polarization vectors;  $k\cdot
x=k_0x_0 -{\bf  k}{\bf  x}$,  $k_0=\omega/c$,  ${\bf  k}=k_0{\bf  n}$,
$\omega$  is the electro-magnetic frequency;  ${\bf n}$ is the guiding
vector of wave front, the weight function $\Phi(x,t)$ is \cite{Kot85}
\begin{equation}
\label{f13}                               
\displaystyle
\Phi=e^{-[(1-\gamma) k\cdot x -\beta k_0 (n_1 x_0 - x_1)]/\gamma}.
\end{equation}
\begin{sloppypar}
\noindent Let us put the function $\Phi$ into  the  set  of  equations
(\ref{condition}),       taking        into        account        that
$\Phi\phi=exp[-i(k_0x_0-{\bf  kx}-\beta k_0 (n_1x_0-x_1))/\gamma]$,  \
$\partial_1\Phi\phi=-i(-k_1+\beta     k_0)\Phi\phi/     \gamma$,     \
$\partial_2\Phi\phi=-i(-k_2)\Phi\phi/\gamma$,       \      $\partial_3
\Phi\phi=-i(-k_3)\Phi\phi/\gamma$,                                   \
$(\partial_0+\beta\partial_1)\Phi\phi=  -ik_0\gamma\Phi\phi$,  \ ${\bf
k}={\bf n}k_0$.  As a result we have the eight algebraic equations  of
the  second order for finding the five unknown group parameters $a,  \
e_{23}, \ e_{32}, \ h_{23}, \ h_{32}$:
\end{sloppypar}
\begin{equation}
\label{condition1}
\begin{array}{rrrr}
-n_2m_3(ah_{23}) & -n_3m_2(ah_{32}) & -(n_2l_2+n_3l_3)(a) +
                                    & l_1(-n_1+\beta)=0;            \\
-n_2l_3(ae_{23}) & -n_3l_2(ae_{32}) & -(n_2m_2+n_3l_3)(a) +
                                    & m_1(-n_1+\beta)=0;            \\
-n_2l_2(ae_{32}) & +n_3l_3(ae_{23}) & -(n_2m_3-n_3m_2)(a) -
                                    & \gamma l_1=0;                 \\
-(-n_1+\beta)l_2(ae_{32})  & -\gamma m_3(ah_{23}) & -[(-n_1+\beta)m_3 
+\gamma l_2](a)  -                  & m_1n_3=0;                     \\
+(-n_1+\beta)l_3(ae_{23})  & -\gamma m_2(ah_{32}) & -[(-n_1+\beta)m_2 
+\gamma l_3](a)  +                  & m_1n_2=0;                     \\
-n_2m_2(ah_{32}) & +n_3m_3(ah_{23}) & -(n_2l_3-n_3l_2)(a) +
                                    & \gamma m_1=0;                 \\
-(-n_1+\beta)m_2(ah_{32})  & +\gamma l_3(ae_{23}) & -[(-n_1+\beta)l_3 
-\gamma m_2](a)  -                  & l_1n_3=0;                     \\
+(-n_1+\beta)m_3(ah_{23})  & +\gamma l_2(ae_{32}) & +[(-n_1+\beta)l_2 
+\gamma m_3](a)  +                  & l_1n_2=0.
\end{array}
\end{equation}
By analogy with relativistic theory we will find the solution  of  the 
set (\ref{condition1}) from the requirements ${\bf E}{\bf H}'=0    \to 
{\bf E}{\bf H}=0$, \ ${\bf E}'^2-{\bf H}'^2=0 \to  {\bf E}^2-{\bf H}^2
=0$ in the relations:
\begin{equation} 
\label{ff}
\begin{array}{c}
{\bf E}'{\bf H}'= \Phi^2{\bf E}{\bf H}+                             \\
+\Phi^2(a^2-1)(E_2H_2+E_3H_3)+\Phi^2a^2[(e_{23}+e_{32})E_2E_3+      \\
(h_{23}+h_{32})H_2H_3++e_{32}h_{32}E_2H_2+e_{23}h_{23}E_3H_3]=      \\
\vspace{2mm}
\Phi^2(1-a^2-a^2e_{23}h_{23})E_1H_1=0;                              \\
{\bf E}'^2-{\bf H}'^2=\Phi^2({\bf E}^2-{\bf H}^2)+                  \\
+\Phi^2(a^2-1)({E_2}^2+{E_3}^3-{H_2}^2-{H_3}^2)+ 
\Phi^2a^2[2(h_{23}-e_{23})E_2H_3+                                   \\
2(h_{32}-e_{23})E_3H_2++{h_{32}}^2{H_2}^2+{h_{23}}^2{H_3}^2-
{e_{32}}^2{E_2}^2-{e_{23}}^2{E_3}^2]=                               \\
\Phi^2(1-a^2+a^2{h_{23}}^2)({E_1}^2-{H_1}^2)=0.
\end{array}
\end{equation}
These relations will be true  if  the  group  parameters  possess  the
following properties:
\begin{equation}
\label{fff}
\begin{array}{ll}
\vspace{1mm}
e_{23}=-e_{32}; \ h_{23}=-h_{32}; & e_{23}=h_{32}; \ e_{32}=h_{23}; \\
\vspace{1mm}
e_{23}h_{23}=(1-a^2)/a^2;         & {h_{23}}^2=(a^2-1)/a^2;         \\
h_{23}=-\sqrt{a^2-1}/a;           & e_{23}=+\sqrt{a^2-1}/a.
\end{array}
\end{equation}
Putting (\ref{fff}) into the set (\ref{condition1}),  we  obtain  four
equations for finding the unknown parameter $a$:
\begin{equation}
\label{ffff}
\begin{array}{rrrr}
\vspace{1mm}
\sqrt{a^2-1}             & - n_1a & + n_1 - \beta       & =0;       \\
\vspace{1mm}
n_1\sqrt{a^2-1}          & - a    & + \gamma            & =0;       \\
\vspace{1mm}
\gamma\sqrt{a^2-1}       & -(-n_1+\beta)a & - n_1       & =0;       \\
(-n_1+\beta)\sqrt{a^2-1} & - \gamma a     & + 1         & =0.
\end{array}
\end{equation}
From this and the set (\ref{fff}) we have $\sqrt{a^2-1}=[n_1(\gamma-1)+
\beta]/(1-{n_1}^2)$ and
\begin{equation}
\label{f14}
\begin{array}{c}
\vspace{1mm}
a=[n_1 (\beta - n_1 ) +\gamma]/[1-{n_1}^2];                         \\
\vspace{1mm}
e_{23}=+[n_1(\gamma-1)+\beta]/[n_1(\beta-n_1)+\gamma];              \\
h_{23}=-[n_1(\gamma-1)+\beta]/[n_1(\beta-n_1)+\gamma].
\end{array}
\end{equation}
As a result the formulas (\ref{f8}) take the form:
\begin{equation}
\label{f17}
\begin{array}{l}
\vspace{1mm}
\displaystyle
{E_1}'=\Phi E_1;                                                    \\
\vspace{1mm}
\displaystyle
{E_2}'=\Phi\frac{[n_1(\beta-n_1)+\gamma]E_2-[n_1(\gamma-1)+\beta]H_3}
{1-{n_1}^2};                                                        \\
\vspace{2mm}
\displaystyle
{E_3}'=\Phi\frac{[n_1(\beta-n_1)+\gamma]E_3+[n_1(\gamma-1)+\beta]H_2}
{1-{n_1}^2};
\vspace{1mm}                                                        \\
{H_1}'=\Phi H_1;                                                    \\
\vspace{1mm}
\displaystyle
{H_2}'=\Phi\frac{[n_1(\beta-n_1)+\gamma]H_2+[n_1(\gamma-1)+\beta]E_3}
{1-{n_1}^2};                                                        \\
\displaystyle
{H_3}'=\Phi\frac{[n_1(\beta-n_1)+\gamma]H_3-[n_1(\gamma-1)+\beta]E_2}
{1-{n_1}^2}.
\end{array}
\end{equation}
(For comparison,   in   relativistic   theory   we   have    $\Phi=1$,
$a=1/\sqrt{1-\beta^2}$,     $e_{23}=-e_{32}=h_{32}=\beta$,     $h_{23}
=-h_{32}=e_{32}=-\beta$).  It can be shown  that  the  transformations
(\ref{f17})  form  a  group  by  virtue  of  the  Galilei  theorem  of
velocities addition $\beta"=\beta + \gamma\beta'$  and  transformation
properties of the guiding cosines of a wave front
\begin{equation}
\label{f18}
{n_1}'=(n_1-\beta)/\gamma; \  {n_2}'=n_2/\gamma; \  {n_3}'=n_3/\gamma.
\end{equation}                                                       
In this case the group parameters and the weight function $\Phi$  have
the following transformation properties:
\begin{equation}
\label{f19}
\begin{array}{c}
\vspace{0.5mm}
\displaystyle
{e_{23}}''=({e_{23}}'+e_{23})/(1+{e_{23}}'e_{23});                  \\
\vspace{0.5mm}
{e_{32}}''=({e_{32}}'+e_{32})/(1+{e_{32}}'e_{32});                  \\
\vspace{0.5mm}
\displaystyle
{h_{23}}''=({h_{23}}'+h_{23})/(1+{h_{23}}'h_{23});                  \\
{h_{32}}''=({h_{32}}'+h_{32})/(1+{h_{32}}'h_{32}); 
\end{array}
\end{equation}
\begin{equation}
\label{f21}
a''=a'a(1+{e_{23}}'e_{23})=a'a(1+{h_{23}}'h_{23})=\ldots;
\end{equation}
\begin{equation}
\label{f20}
\Phi''=\Phi'\Phi,
\end{equation}
which correspond to the matrix law  of  multiplying  the  matrices  of
field  transformations  in  going  to  primed variables.  The formulas
(\ref{f19})  are  completely  analogous  to   the   formula   of   the
relativistic   theorem  of  velocities  addition  and  also  arise  in
multiplying the matrices from the Lorentz group.  From this and due to
the  presence  of the local weight function $\Phi(x,t)$ (\ref{f13}) it
can be concluded  that  the  matrices  of  the  field  transformations
(\ref{f17})  form  the  projective representation of the Lorentz group
\cite{Ham66} on the solutions as plane waves of  D'Alembert  equation.
It   is  similar  to  the  projective  Galilei  group  representations
realizing on the solutions as plane waves of Schr\"odinger equation in
Quantum   theory  \cite{Hag72},  \cite{Nie72},  and  it  is  also  the
unexpected circumstance in the symmetry theory of  Maxwell  equations.
But the similar cases are known.
\par For  example,  as  early as 1909,  Cunningham \cite{Cun09} showed
that the inversion group $I$:  ${x^\nu}'=x^\nu/x^2 \ (x^2={x^0}^2-{\bf
x}^2, \  x^0=ct$) in Minkowski space induces the electromagnetic field
transformations,  which can be described through a matrix  of  Lorentz
group    representation    $D(L)$    with    the    local   parameters
$\beta=2x^0r/({x^0}^2+r^2)$,                                         \
$\sqrt{1-\beta^2}=({x^0}^2-r^2)/({x^0}^2+r^2)$:
\begin{equation}
\label{cun}
\phi'_p(x')=x^4 D_{pq}(L)\phi_q (x),
\end{equation}
where $r^2=x^2+y^2+z^2$,  \ $x^4=({x^0}^2-{\bf x}^2)^2$;  $\phi_p  \in
({\bf E}, {\bf H})$, \ $p,q=1,2,\cdots,6$.
\par In 1970 Isham,  Salam and Strathdee generalized this  result  for
the special conformal group $C_4$: $x'^\mu=(x^\mu-a^\mu x^2)/\sigma, \
\sigma= (1-2a\cdot  x+a^2x^2)$  and  fields  of  different   conformal
dimensions \cite{Ish70}. We  write  their  result,  as  applied to $4$
-potential $(A^0, {\bf A})$ of electromagnetic field
\begin{equation}
\label{ish}
\phi'_p(x')=
\begin{array}{|c|}
det\partial x'/\partial x\end{array} \ ^{l/4}
{\rm D}_{pq}(L)\phi_q (x)=\sigma {\rm D}_{pq}(L)\phi_q,
\end{equation}
where $\phi \in (A^0,  {\bf A})$,  \ $p,q=0,1,2,3$,  \ $l=-1$  is  the
conformal dimension of $4$ -potential.  (The  conformal  dimension  is
equal  $-2$  for the electromagnetic fields ${\bf E},  {\bf H}$ in the
case of Cunningham).
\par The  same  result  ($l=-1$)  may  be  also received by the method
\cite{Kot96}.
\par Now  let  us turn to the our case.  One can see that the formulas
(\ref{f17}) for electromagnetic field transformations may  be  written
as:
\begin{equation}
\label{kot}
\phi'_p(x')=\Psi(x) D_{pq}(L)\phi_q (x)
\end{equation}
with the weight function  $\Psi$  and  the  group  parameters  $a$,  \
$e_{23},  \  h_{23}$ from the expressions (\ref{f13}) and (\ref{f14}).
We have an analogy with the result  \cite{Cun09}  exception  that  the
weight  function  $x^4$  in the expression (\ref{cun}) has a universal
character,  but in the formula  (\ref{kot})  the  weight  function  is
determined  by  a concrete solution of Maxwell equations;  besides the
Lorentz,  Inversion groups are the subgroups of  the  conformal  group
$O(2,4)$  and  the  Lorentz,  Galilei  groups are the subgroups of the
$O(2,5)$ group \cite{Kuz92}.
\par In sum, the formulas (\ref{f17}) and (\ref{kot}) reflect the fact
that  the  electromagnetic  fields  may  be  classified  through their
transformation properties under the Lorentz group.  The given property
of   electromagnetic   field   was  also  investigated  in  the  paper
\cite{Kot96}  as  evidence  of  the  symmetry  of  Maxwell  equations,
conditioned  by  existence  of  the  second order commutational ratios
$[\Box [\Box,{\it H}_k]]=0$,  where ${\it H}_k$ are the generators  of
Galilei group, $k=1,2,3$, $\Box$ is D'Alembert operator.
\par So,  according to the result received,  we may say that the  well
known  Lorentz-symmetry  of  Maxwell  equation is realized on the true
representations of  Lorentz  group  \cite{Ham66};  on  the  projective
representations   of  Lorentz  group  the  Galilei-symmetry  of  these
equations is realized.
\par We also note that if the guiding vector is ${\bf n}=(1,0,0)$, the
field transformations is reduced to formulas
\begin{equation}
\label{adf17}
\begin{array}{l}
\vspace{1mm}
\displaystyle  E_1'=E_1=0;                                          \\
\vspace{1mm}
\displaystyle
E_2'=\frac{[(1-\beta+0.5\beta^2)E_2-\beta(1-0.5\beta)H_3]}{1-\beta};\\
\vspace{2mm}
\displaystyle
E_3'=\frac{[(1-\beta+0.5\beta^2)E_3+\beta(1-0.5\beta)H_2]}{1-\beta};\\
\displaystyle
\vspace{1mm}    H_1'=H_1=0;                                         \\
\vspace{1mm}
\displaystyle
H_2'=\frac{[(1-\beta+0.5\beta^2)H_2+\beta(1-0.5\beta)E_3]}{1-\beta};\\
\displaystyle
H_3'=\frac{[(1-\beta+0.5\beta^2)H_3-\beta(1-0.5\beta)E_2]}{1-\beta},
\end{array}
\end{equation}
were $\Phi=1$,  \  $\gamma=(1-\beta)$,  \  $a=(1-\beta+0.5\beta^2)/(1-
\beta)$, \ $e_{23}=\beta(1-0.5\beta)/(1-\beta+0.5\beta^2)$.
\par Considerable  for the theory is the question on the invariants of
the space-time  and  field  transformations.  Let  us  write  them  in
comparison with the relativistic case.

\begin{equation}
\label{f22}
\begin{array}{ll}
The \ Galilei \ group \ {\it G}_1 & The \ Lorentz \ group \ {\it L}_1 \\
{}                                                                    &
{}                                                                    \\
\vspace{1mm}
{c'}^2{t'}^2-{\bf x}'^2=c^2t^2-{\bf x}^2=0;                           &
{c'}^2{t'}^2-{\bf x}'^2=c^2t^2-{\bf x}^2=s^2;                         \\
\vspace{1mm}
t'=t;   \ c<\infty;                                                   &
c'=c;   \ t<\infty;                                                   \\
\vspace{1mm}
{{k_0}'}^2 - {{\bf k}'}^2={k_0}^2-{\bf k}^2=0;                        &
{{k_0}'}^2 - {{\bf k}'}^2={k_0}^2-{\bf k}^2=0;                        \\
\vspace{1mm}
{k_0}'=k_0; {{\bf k}'}^2={\bf k}^2;                                   &
-                                                                     \\
{{\bf n}'}^2={\bf n}^2=1;                                             &
{{\bf n}'}^2={\bf n}^2=1;                                             \\
\vspace{1mm}
{n_1}'{x_0}'-{x_1}'=n_1x_0-x_1;                                       &
-                                                                     \\
\vspace{1mm}
{\bf E}'{\bf H}'=\Phi^2 {\bf E}{\bf H}=0;                             &
{\bf E}'{\bf H}'={\bf E}{\bf H}=0;                                    \\
\vspace{1mm}
{{\bf E}'}^2-{{\bf H}'}^2=\Phi^2 ({{\bf E}'}^2-{{\bf H}'}^2)=0;       & 
{{\bf E}'}^2-{{\bf H}'}^2={{\bf E}'}^2-{{\bf H}'}^2=0;                \\
\vspace{1mm}
{E_1}'=\Phi E_1; \ {H_1}'=\Phi H_1;                                   &
{E_1}'=E_1;      \ {H_1}'=H_1;                                        \\
\Box'=[(\partial_0+\beta\partial_1)^2/\gamma^2-\triangle];            &
\Box'=\Box.
\end{array}
\end{equation}
It is  not  difficult  to  see  that  the  part  of Galilei-invariants
coincides with Lorentz-invariants.  However, the distinctions are also
observed.  They  are  caused  by  the  distinction in transformational
properties  of  the  appropriate  values.  For  example,  formulas  of
transformation  of the zero vector components $(k_0,  {\bf k})$ in the
Galilean and relativistic variants of the theory are different:
\begin{equation}
\label{f23}
\begin{array}{ll}
The \ Galilei \ group \ {\it G}_1 & The \ Lorentz \ group \ {\it L}_1 \\
{}                                                                    &
{}                                                                    \\
\vspace{1mm}
{k_0}'=k_0;                                                           &
\vspace{1mm}                                                    
{k^0}'=(k^0-\beta k^1)/\sqrt{1-\beta^2};                              \\ 
\vspace{1mm}                                                    
{k_1}'=(k_1-\beta k_0)/\gamma;                                        &
{k^1}'=(k^1-\beta k^0)/\sqrt{1-\beta^2};                              \\
{k_2}'=k_2/\gamma; \ {k_3}'=k_3/\gamma;                               &
{k^2}'=k^2; \  {k^3}'=k^3.
\end{array}
\end{equation}                                                       
They differently describe the Doppler effect in electrodynamics.
\par Let us consider the approximation of the  small  speeds.  At  the
small speeds,  when $\beta<<1$, for both the relativistic and Galilean
cases the common formulas of field transformations are realized:
\begin{equation}
\label{f24}
\begin{array}{cc}
{\bf E}'={\bf E} + {\bf V}{\rm x} {\bf H}/c;                           &
{\bf H}'={\bf H} - {\bf V}{\rm x} {\bf E}/c
\end{array}
\end{equation}
They are  known  as the Galilean limit.  In the present work this term
has a conditional character,  as the formulas are the limit  both  for
the  relativistic and nonrelativistic field variables transformations.
The  situation  here  is  analogous  to  the  case  with   independent
variables:
\begin{equation}
\label{f25}
x'=x-Vt, \ y'=y, \ z'=z, \ t'=t, \ c'=c.
\end{equation}                                                       
These transformations being neither Galilean, nor relativistic are the
limit ratios both for the Galilei and relativistic cases. 
\par In  the  approximation of ultra high velocities,  when $\beta \to
1$,  the  electromagnetic  fields   ${\bf   E}=(0,1,0)\phi,   \   {\bf
H}=(0,0,1)\phi$  with ${\bf n}=(1,0,0)$,  \ $\phi=exp(-ik\cdot x)$ are
transformed as
\begin{equation}
\label{g}
\phi'=(1-\beta)\phi; \ \phi'=\sqrt{(1-\beta)/(1+\beta)}.
\end{equation}
in the  Galilean and relativistic cases respectively.  Their principle
distinction consists in the absence of real fields in the relativistic
case if $\beta>1$.
\par We also note,  that in the relativistic theory the transformation
parameters  of the field variables do not depend on the kind of field.
Owing to this fact these transformations have  global  nature  in  the
space  of solutions.  In Galilean case,  the transformation parameters
and the weight functions in space of solutions are determined  by  the
concrete  field.  In this sense the Galilean field transformations are
non linear. Just this circumstance is the reason why Galilean symmetry
and  the  associated  concept of the Newton time in electrodynamics so
long slipped off from  researchers,  as  for  all  the  cases,  except
\cite{Kot85}  -  \cite{Kot91},  the  field  transformations were being
searched in the class of the linear transformations.

     These are the mathematical consequences from the introductions of
the Newton time in electrodynamics.  It is natural the question of the
physical consequences arises as well. We consider only some of them.

     If we admit in the framework of Galilei transformations, that the
speed  of  light  is equal $3.10^{10} cm/s$ with respect to an emitter
(Ritz hypothesis \cite{Pau47}),  according to the Galilean theorem  of
velocities  addition  the speed of light should depend on the velocity
of movement of the emitter.  However such effects have not been  found
in   the  specially  carried  out  experiments.  (Bonch-Bruevitch  and
Molchanov,  1956;  Alv\"ager,  Nilsson and Kjellman,  1963;  James and
Sternberg,  1963;  Sadeh,  1963;  Filippas and Fox,  1964;  Alv\"ager,
Farley,  Kjellman and Wallin,  1964; Babcock and Bergman, 1964). It is
possible  to  overcome the marked difficulty,  when we assume that the
speed of light is equal  $3.10^{10}  cm/s$  not  with  respect  to  an
emitter,  but with respect to an observer \cite{Kot85}. For this case,
the relationship between the speed of light and the  velocity  of  the
emitter is eliminated,  but it is necessary to assume that any emitter
beforehand knows,  with which the speed it should  radiate  the  light
that  the  speed of this light relative to any observer would have the
required value $3.10^{10} cm/s$.  This hypothesis and the  experiments
connected with it have been discussed in the paper \cite{Kot85}.  Here
we consider an another approach.

\section{The extended Galilei transformations}
\label{3}
The extended Galilei group ${\it G}_{11}$ may be defined as the set of
linear space-time transformations
\begin{equation}
\label{e1}
{\bf x}'={\varrho}(R{\bf x}+{\bf V}t)+{\bf a}, \ t'=t+b,
\end{equation}
where $R$ is the matrix of the 3-dimensional spatial rotations;  ${\bf
V}$  is  the velocity of the inertial reference frame $K'$ relative to
$K$;  ${\varrho}$ is the scale parameter.  The transformations induced
are  defined  in 4-dimensional real space $E^3\bigotimes T^1$ with the
metric
\begin{equation}
\label{e2}
ds'^2=dx'^2+dy'^2+dz'^2={\varrho}^2(dx^2+dy^2+dz^2)={\varrho}^2ds^2
\end{equation}
in the  subspace $E^3\subset E^3\bigotimes T^1$.  The extended Galilei
group  Lie  algebra  is   generated   by   the   set   of   operators:
$p_0=i\partial_t, \ p_k=-i\partial_k$ of the translations group $T_4$;
${\it J}_k={({\bf x}{\rm x}{\bf p})}_k$  of  the  3-spatial  rotations
group   $SO_3$;   ${\it   H}_k=-tp_k$   of   the   the   pure  Galilei
transformations group ${\it G}_3$;  ${\it  D}=-x_kp_k$  of  the  scale
transformation     group     $\triangle_1$,    where    $k=1,2,3;    \
x_{1,2,3}=x,y,z$.  The commutational ratios of  these  generators  are
\cite{Kot86}, \cite{Hag72}, \cite{Nie72}:
\begin{equation}
\label{e3}
\begin{array}{lll}
\vspace{1mm}
\displaystyle [{\it J}_k,p_0]=0; & [{\it H}_k,p_0]=ip_k; & [p_k,p_0]=0;\\
\vspace{1mm}
\displaystyle [{\it J}_k,p_l]=i\epsilon_{klm}p_m; & [{\it H}_k,p_l]=0; & 
[p_k,p_l]=0; \\
\vspace{1mm}
\displaystyle [{\it J}_k,{\it  H}_l]=i\epsilon_{klm}{\it  H}_m;        &  
[{\it  H}_k,{\it H}_l]=0;        & [p_0,{\it D}]=0;                    \\
\vspace{1mm}
\displaystyle [{\it J}_k,{\it  J}_l]=i\epsilon_{klm}{\it  J}_m;        &
[{\it  H}_k,{\it D}]=i{\it H}_k; &                 [p_k,{\it D}]=ip_k; \\
\displaystyle [{\it J}_k,{\it D}]=0, & {} &                            {}
\end{array}  
\end{equation}
where $\epsilon_{klm}$  is  the completely antisymmetrical tensor with
$\epsilon_{123}=1$.  The  Lie  algebra  generators  are  the  symmetry
operators   of   Maxwell   equations   because  of  existence  of  the
commutational ratios $[\Box,p_0]=  [\Box,p_k]=[\Box,{\it  J}_k]=0$,  \
$[\Box[\Box,{\it  H}_k]]=0$,  \ $[\Box[\Box,D]]=0$ \cite{Kot91}.
\par Bearing this in mind, let us choose the parameter $\varrho$ as
\begin{equation}
\label{e4}
\varrho=\gamma^{-1}=1/(1-2\beta R_{kl}s_kn_l+\beta^2)^{1/2}
\end{equation}
with ${\bf s}={\bf V}/V$, \  ${\bf n}={\bf c}/c$, \ $\beta=V/c$.
Then the   extended   Galilei   transformations   (\ref{e1})  will  be
compatible both with the concept of the universal Newton  time  $t'=t$
and  with the postulate of invariance of the speed of light because we
may conclude from the theorem of velocities addition  that  $c'=c$  in
the given  case.  Owing  to  this  fact,  for the particular case when
$R_{kl}=\delta_{kl}$,   \   ${\bf   V}=(V,0,0)$,    \    $b=0$,    the
transformations of space and time are \cite{Kot91}
\begin{equation}
\label{f26}
\begin{array}{c}
\vspace{1mm}
\displaystyle
x'=\frac {x-Vt}{\sqrt{1-2\beta n_x+\beta^2}};                       \
\displaystyle
y'=\frac {y}{\sqrt{1-2\beta n_x+\beta^2}};                          \
\displaystyle
z'=\frac {z}{\sqrt{1-2\beta n_x+\beta^2}}; \ t'=t; \ c'=c,
\end{array}
\end{equation}                              
where group parameters have the following properties:
\begin{equation}
\label{e5}
\begin{array}{c}
\vspace{1mm}
\gamma'=1/\gamma;       \ \beta'=-\beta/\gamma;                      \\
\gamma''=\gamma\gamma'; \ \beta''=\beta+\gamma\beta'.
\end{array}
\end{equation}

\subsection{Invariants of the extended transformations}
\label{3a}
We note invariants of extended transformations induced once again.
\smallskip

\noindent {\bf The speed of light}.
We have    from    the    transformational   properties   of  velocity
$v_x'=(v_x-V)/\gamma, \ v_y'=v_y/\gamma, \ v_z'=v_z/\gamma$:
\begin{equation}
\label{e6}
c'=\sqrt{c_x'^2+c_y'^2+c_z'^2}=\sqrt{(c_x-V)^2+{c_y}^2+{c_z}^2}/
\sqrt{1-2\beta n_x+\beta^2}=c.
\end{equation}
\smallskip

\noindent {\bf The equation of propagation of a spherical wave}. 
Suppose $c'^2t'^2-{\bf x}'^2=0$. Then
\begin{equation}
\label{e7}
c'^2t'^2-{\bf x}'^2=(n_x'^2+n_y'^2+n_z'^2)c'^2t'^2-{\bf x}'^2=
(c^2t^2-{\bf x}^2)/\sqrt{1-2\beta n_x+\beta^2},
\end{equation}
where $x=cn_x t, \ n_x'=(n_x-\beta)/\gamma, \ n_y'=n_y/\gamma, \ n_z'=
n_z/\gamma$. Hence we obtain from $c'^2t'^2-{\bf x'}^2=0$ that
$c^2t^2-{\bf x}^2=0$.
\hfil\break
\smallskip

\noindent {\bf The radius of spherical wave}.
Suppose $R^2=x^2+y^2+z^2=c^2t^2, \ R=ct$. Then
\begin{equation}
\label{e8}
\begin{array}{l}
\displaystyle
R'=\sqrt{x'^2+y'^2+z'^2}=\sqrt{x^2-2Vxt+V^2t^2+y^2+z^2}/\gamma=
\gamma ct/\gamma=R.
\end{array}
\end{equation}
\smallskip

\noindent {\bf The sum of square of guiding cosines}.
Suppose ${\bf n}^2=1$. Then
\begin{equation}
\label{e9}
{\bf n}'^2=(n_x^2-2\beta n_x+\beta^2+{n_y}^2+{n_z}^2)/
(1-2\beta n_x+\beta^2)=1.
\end{equation}
\smallskip

Hence, in  accordance  with  the  principle of relativity the extended
Galilei transformations (\ref{f26}) transform  the  sphere  of  radius
$R'=c't'$  with center $x'=y'=z'=0$ to the sphere with the same radius
$R=ct=c't'=R'$  and  center  $x=y=z=0$.  The  speed  of   light   $c'$
transforms  to  the  same speed of light $c=c'$.  The sum of square of
guiding cosines turns to unit.  We obtain an analogy with SR.  We also
note  that  in  the  framework  of  extended  Galilei  transformations
(\ref{f26}) the consecutive  explanation  of  the  basic  relativistic
experiments is possible.  Let us consider these explanations following
the papers \cite{Kot85},  \cite{Kot91}.

\section{The elements of physical interpretation}
\label{4}

\subsection{Similar to SR interpretation of observations}
\label{4a}

{\bf The Michelson experiment}.  Let the test run with the terrestrial
light  source,  System  K.  Then  by virtue of isotropy of 3-space the
speed of light c will be the same in all directions.  Therefore, for a
terrestrial  observer  immobile with respect to an interferometer, the
interference pattern will  not  be  changed  when  the  interferometer
is rotated.  Then  let the test run with an extraterrestrial source of
light,  System K'.  In this case, due to the transformational property
c'=c -inv the speed of light in the system K, i.e. on the Earth, turns
out to be equal c and will coincide with the speed of light  from  the
terrestrial   source.  Therefore,  experiments  with  terrestrial  and
extraterrestrial sources of light  are  indistinguishable  physically,
and  the  interference  pattern will not be changed in our case as in
SR.
\bigskip

\noindent {\bf  Test  to check independence of the speed of light from
velocity of the light source}.  As in SR,  all these experiments  will
lead to negative results since $c'=c$.
\bigskip

\noindent {\bf  The Fizeau experiment}.  Let us suppose that the speed
of light in moving medium (water) is equal
\begin{equation}
\label{f27}
v'=\frac{c}{n}+V\biggl(\frac{1}{n}-\frac{1}{n^2}\biggl),
\end{equation}
where $n$ is the refractive index,  $V$ is the velocity of water. Then
according  to  the  velocity  addition  theorem  (following  from  the
transformations (\ref{f26}) ) in the laboratory  frame  the  speed  of
light turns out to be
\begin{equation}
\label{f28}
v=v'\sqrt{1-2\beta n_1 +\beta^2}+V=(1-\beta)\biggl[\frac{c}{n}+V\biggl
(\frac{1}{n}-\frac{1}{n^2}\biggl)\biggl]\approx\frac{c}{n}+V\biggl(1-
\frac{1}{n^2}\biggl)
\end{equation}
at $n_1=1$ in agreement with the experiment. Hence, the explanation of
the Fizeau experiment is conventional.  For example,  as compared with
the formula (\ref{f27}) we have $v'=c/n$ in SR, in classical physics -
$v'=(c/n-V/n^2)$ \cite{Pau47}.
\bigskip

\noindent {\bf Aberration of light}.  Let $K'$ be the Sun,  $K$ be the
Earth.  For  explanation  the effect,  as in SR,  we assume that ${\bf
V}=(V,0,0)$ is the velocity of the Sun and that the light propagate in
the direction ${\bf n}'=(0,1,0)$ in frame $K'$.  Then we have from the
transformational properties of the guiding vector  ${\bf  n}$  of  the
speed of light
\begin{equation}
\label{f29}
\begin{array}{c}
\vspace{2mm}
\displaystyle
n_1'=(n_1-\beta)/\gamma=0; \ n_2'=n_2/\gamma=1; \ n_3'=n_3/\gamma=0; \
\displaystyle
\gamma=\sqrt{1-2\beta n_1+\beta^2};                                  \\
\displaystyle
n_1=cos\theta=\beta; \ n_2=sin\theta=\gamma; \ n_3=0;                \
\displaystyle
\gamma=\sqrt{1-\beta^2}.
\end{array}
\end{equation}
Since the half of aberration angle is equal to $\alpha=(\pi/2-\theta)$
we   obtain   $sin\alpha=\beta$,  $\alpha\approx\beta=V/c=10^{-4}$  in
agreement with  the  experiment.  In  more  general  case  with  ${\bf
n}'=(cos\theta',  sin\theta',  0)$  the angle of view on a star may be
found  from  the  expression  $tg\theta'=sin\theta/(cos\theta-\beta)$.
This     formula     coincides     with     the    relativistic    one
$tg\theta'=sin\theta\cdot\sqrt{1-\beta^2}/(cos\theta-\beta)$
\cite{Pau47} with an accuracy $\beta^2$ \cite{Kot85}.
\bigskip

\noindent {\bf The Doppler effect}. Suppose that an emitter moves with
a velocity $V$ along the $x$ -axis relative to an observer $K$. Let us
attach the frame $K'$ to  the  emitter,  and  take  into  account  the
formulas   ${k_0}'=\gamma k_0$,   \   ${k_1}'=(k_1-\beta  k_0)$,     \
${k_2}'=k_2$, \ ${k_3}'=k_3$ (instead of the the formulas (\ref{f23})).
Then we have:
\begin{equation}
\label{f30}
{k_0}'=\gamma k_0 \to (\omega_0'/c')=
\sqrt{1-2\beta n_1+\beta^2} \ (\omega/c); \ c'=c.
\end{equation}
From this we obtain
\begin{equation}
\label{f31}
\omega=\omega_0/{\sqrt{1-2\beta n_1+\beta^2}}; \ 
\lambda=\lambda_0 \sqrt{1-2\beta n_1+\beta^2},
\end{equation}
where ${\omega_0}'=\omega_0$  is  the proper frequency and $\lambda_0$
is  the  proper  wavelength  of  radiation;  $\omega$,  $\lambda$  are
respectively  the  frequency and wavelength,  measured experimentally.
From the results (\ref{f31}) we have the following frequencies for the
longitudinal $(n_1\approx1)$ and transversal $(n_1=0)$ Doppler effect:
\begin{equation}
\label{adf31}
\begin{array}{cc}
\omega_{\parallel}=\omega_0[1+\beta n_1-\beta^2(1-3{n_1}^2)/2];      &
\omega_{\perp}=\omega_0[1-\beta^2/2-\beta^4/8].
\end{array}
\end{equation}
For comparison, in the relativistic case the analogous formulas are:
\begin{equation}
\label{addf31}
\begin{array}{cc}
\vspace{2mm}
\omega=\omega_0\sqrt{1-\beta^2}/(1-\beta n_1);                       &
\lambda=\lambda_0(1-\beta n_1)/\sqrt{1-\beta^2};                     \\
\displaystyle
\omega_{\parallel}=\omega_0[1+\beta n_1-\beta^2(1-2{n_1}^2)/2];      &
\omega_{\perp}=\omega_0[1-\beta^2/2-3\beta^4/8].
\end{array}
\end{equation}
These formulas    coincide   with   the   formulas   (\ref{f31})   and
(\ref{adf31}) for the longitudinal and the transverse Doppler  effects
in Galilean case with an accuracy $\beta^2$ and $\beta^4$ \cite{Kot85}.
\bigskip

\subsection{Distinction from SR interpretation of observations}
\label{4aa}
We note some of them.
\bigskip

\noindent {\bf  The  twin  paradox}.  In  Galilean  approach  the twin
paradox do not exist because of universal character of time.
\bigskip

\noindent {\bf   Tests   to   measure   lifetimes  of  fast  nonstable
particles}. In the framework  of  transformations  (\ref{f26})  it  is
necessary to   admit,   that   instead  of  the  effect  of  the  time
retardation   the  movement  with  superrelativistic  velocity  should
exist.  For example,  the velocity of atmospheric $\mu$ -mesons should
be equal  to  $\sim  6\cdot  10^6/2\cdot  10^{-6}=3\cdot  10^{12}cm/s$
\cite{Kot85}.
\bigskip

\begin{sloppypar}
\noindent {\bf Superrelativistic  objects  with  real  mass}.  
In the  relativistic  physics  such  objects  can  not  exist.  In the
Galilean case their existence  is  not  forbidden.  Besides  the  fast
nonstable  particles,  such  objects may be space objects with a large
value of the redshift
\end{sloppypar}
\begin{equation}
\label{f33}
z=\frac{\lambda-\lambda_0}{\lambda_0}=\sqrt{1-2\beta n_1+\beta^2} \ -1
\end{equation}
In the case of a radial movement the  parameter  of  the  redshift  is
$z\simeq\beta(\beta-n_1)/(\beta+1)\simeq\beta$  when  $n_1\simeq  -1$,
and for $z>1$ the superrelativistic velocity of object will  be  equal
$v_{\parallel}\simeq zc>c$ \cite{Kot91}.  For the transversal movement
when $n_1=0$,  the redshift is $z=\sqrt{1+\beta^2}-1$ and the velocity
of  \  object  will  be equal $v_{\perp}=\sqrt{z(z+2)} \ c$.  Then the
superrelativistic movements may be if $z>\sqrt2-1=0.414$.  The  quasar
NRAO  140 with redshift z=1.258 may be an object of such type.  In the
framework of Galilei approach the radial calculated velocity  of  this
object may be equal $v_{\parallel}\simeq zc=1.258 c$;  the transversal
calculated     velocity     may     be     equal      $v_{\perp}\simeq
\sqrt{z(z+2)}=2.02c$.  The surprising thing is the fact that the value
$2.02c$ is close to the lower limit of the superrelativistic expansion
velocity   of   QSO  NRAO  140  in  the  framework  of  the  Friedmann
cosmological model:  from $3c$ to $10c$,  depending  upon  assumptions
\cite{Six81}.
\par Let  us  also  note,  that the considered example is not the only
case when a quasar has the  value  of  redshift  $z>1$.  According  to
\cite{Ku{}80},  the  number  of  these  quasars  is great enough,  for
example, the quasars LB8796  z=1.320,  PKS  z=2.170,  PHL957  z=2.690,
OQ172 z=3.530, etc.  It  is  known more than 100 objects with redshift
$z>3$ too \cite{Uni90},  among which the quasar Q1158+4635 with z=4.73
is present.  From standpoint of the work all these  objects  may  move
with   superrelativistic   velocities   $v_{\parallel}\simeq  zc$  and
superrelativistic motion is an ordinary  phenomenon  in  astrophysics.
Hubble's  distance for these objects is equal $D_H\simeq zc/H$ and may
be connected with Friedmann's distance  (Robertson-Walker  metric)  by
the relation: $D_F=D_H\{q_0+ (q_0-1)[(1+2q_0z)^{1/2}-1]/z\}/{q_0}^2(1+
z)$,  where $H$ is the Hubble  constant,  $q_0$  is  the  deceleration
parameter   \cite{Coh71}.   For   the   case   of   $q_0=1$   we  have
$D_F=D_H/(1+z)=zc/H(1+z)$ and the horizon $c/H$,  which is  absent  in
the Galilean approach,  occurs in Friedmann's model.  If $q\to 0$, the
Friedmann's distance $D_F\to z^2c/H(1+z)$ is  close  to  the  Galilean
value  $D_H\simeq  zc/H$  for  the  great  values  of  the  red  shift
parameters $z\gg 1$.
\bigskip

\noindent {\bf The limiting  speed  of  propagation  of interactions}.
According  to  the addition velocities theorem,  we have the following
value $v$ for movement along of the $x$ - axis at ${\bf n}= (1,0,0)$:
\begin{equation}
\label{f32}
v=\gamma v'+V=v'+V(1-v'/c).
\end{equation}
Here the value $v'=c$ has the invariance property only for the case of
such phenomenon as the propagation of the light.  If $v'\neq  c$,  the
velocity  $v$  may  be  any  amount  great  by  appropriate value $V$.
Therefore the limiting speed of propagation of all interactions in the
approach (\ref{f26}) do not exist.  The interactions having the nature
other  than  the  electromagnetic  one  may  propagate  with  velocity
distinct from the value $c=3\cdot 10^{10}cm/s$.
\bigskip

\noindent  {\bf The point-behaviour  of  elementary  particles}.
The elementary  particles should be points in the relativistic physics
because of the finite speed of propagation  of  all  interactions.  In
Galilean  physics  this  requirement  may  be  removed  because of the
absence of the limiting speed of interactions propagation.

\section{Conclusion}
\label{5}
It is  shown  that  in  the frame of the generalized symmetry approach
\cite{Kot96} Maxwell equations are invariant with respect  to  Galilei
transformations  and admit the existence of the universal Newton time.
In the case of the extended Galilei transformations the  postulate  of
the  universal  Newton time may be made compatible with the concept of
invariance of the speed of light.  The  Galilei  symmetry  of  Maxwell
equation  means  that the Galilei relativity principle may be realized
not  only  in  the  classical   mechanics   but   in   the   classical
electrodynamics too.

\end{document}